\documentclass[conference]{IEEEtran}

\usepackage{amsmath}
\usepackage{amsfonts}
\usepackage{amssymb}
\usepackage{geometry}
\usepackage{booktabs}
\usepackage{transparent}
\usepackage{acronym}
\usepackage{graphicx}
\usepackage{adjustbox}
\usepackage{array}
\usepackage{lscape}
\usepackage{pifont}
\usepackage{algorithm, algpseudocode}
\usepackage{fancyhdr, geometry}
\usepackage{balance}
\usepackage{tikz}

\usetikzlibrary{shapes.geometric, arrows}
\newcommand{\xmark}{\ding{55}} 

\acrodef{RBS}{Range-Based Sharding Protocol}
\acrodefindefinite{RBS}{an}{a}
\acrodef{PoW}{Proof of Work}
\acrodef{PoS}{Proof of Stake}
\acrodef{PoA}{Proof of Authority}
\acrodef{PBFT}{Practical Byzantine Fault Tolerance}
\acrodef{BFT}{Byzantine Fault Tolerance}
\acrodef{2PC}[2PC]{Two-Phase Commit}
\acrodef{DoS}{Denial-of-Service}
\begin{document}

\bstctlcite{IEEEexample:BSTcontrol}

\title{A Range-Based Sharding (RBS) Protocol for Scalable Enterprise Blockchain}

\author{M.Z. Haider,M. Dias de Assunção, Kaiwen Zhang \\
$^{1}$Department of Software Engineering, ÉTS Montréal, Universite de Quebec, Canada}

\maketitle

\begin{abstract}
Blockchain technology offers decentralization and security but struggles with scalability, particularly in enterprise settings where efficiency and controlled access are paramount. Sharding is a promising solution for private blockchains, yet existing approaches face challenges in coordinating shards, ensuring fault tolerance with limited nodes, and minimizing the high overhead of consensus mechanisms like PBFT. This paper proposes the Range-Based Sharding (RBS) Protocol, a novel sharding mechanism tailored for enterprise blockchains, implemented on Quorum. Unlike traditional sharding models such as OmniLedger and non-sharding Corda framework, RBS employs a commit-reveal scheme for secure and unbiased shard allocation, ensuring fair validator distribution while reducing cross-shard transaction delays. Our approach enhances scalability by balancing computational loads across shards, reducing consensus overhead, and improving parallel transaction execution. Experimental evaluations demonstrate that RBS achieves significantly higher throughput and lower latency compared to existing enterprise sharding frameworks, making it a viable and efficient solution for large-scale blockchain deployments.
\end{abstract}


\acresetall

\section{Introduction}

Blockchain technology offers decentralization, security, and transparency, making it attractive for various applications, including finance and supply chain management \cite{javaid2021blockchain}. While public blockchains like Bitcoin and Ethereum are well known, permissioned blockchains \cite{kumar2022permissioned} provide a more controlled environment suitable for enterprise needs, offering better privacy and efficiency. However, scalability remains a key challenge for permissioned blockchains as enterprises expand their networks \cite{helliar2020permissionless, khan2021systematic}. Traditional scalability solutions, such as increasing block size or reducing block time, often fall short in permissioned settings where a balance between security, scalability, and efficiency is crucial \cite{kaur2020scalability, pawar2021study}. Blockchain sharding has emerged as a promising technique to address this issue by dividing the network into smaller, parallel segments or ``shards'' \cite{yang2020review}. Each shard processes transactions independently, increasing throughput. Although public blockchains have adopted sharding~\cite{khan2021systematic, xu2023survey}, adapting it to the unique needs of permissioned blockchains remains an area ripe for exploration \cite{kim2018survey, sanka2021systematic, herrera2016privacy}. Traditional approaches to improving scalability in blockchain systems \cite{wang2019sok} \cite{chauhan2018blockchain}, such as increasing block size or reducing block generation time, have proven insufficient, particularly in the context of permissioned blockchains \cite{helliar2020permissionless}\cite{zhou2020solutions}. Such solutions are designed with public blockchains in mind, where the focus is often on maximizing decentralization rather than optimizing performance \cite{zhou2020solutions}\cite{kaur2020scalability}\cite{pawar2021study}. However, adapting sharding to permissioned blockchains presents unique challenges. These include managing the consensus overhead of traditional mechanisms like PBFT or IBFT \cite{chen2023reputation} and some DAG based solutions\cite{noreen2023advanced}, ensuring atomicity and consistency in cross-shard transactions, and addressing fault tolerance within shards~\cite{khan2021systematic, xu2023survey}. Additionally, dynamic shard reconfiguration to handle varying workloads without disruption, and maintaining the privacy and access control requirements of permissioned environments, further complicate sharding adoption\cite{kumar2022permissioned}. These challenges underscore the need for tailored solutions to enable efficient sharding in permissioned blockchains\cite{chauhan2018blockchain}.

This paper introduces a novel approach using \ac{RBS},  designed for enterprise blockchain environments  where traditional sharding such as OmniLedger, Corda, Near protocol is not effective, specifically implemented on the Quorum platform, a permissioned variant of Ethereum that is widely used in enterprise settings. Our method incorporates a commit-reveal scheme for secure randomness generation in shard allocation and consensus, enabling parallel transaction processing while maintaining network integrity.

Our contributions include:

\begin{enumerate}
\item \textbf{Optimized Range-Based Sharding:} Utilizing range-based partitioning with adaptive load balancing, our mechanism enhances transaction throughput while minimizing cross-shard communication overhead.
\item \textbf{Efficient and Secure Cross-Shard Transactions:} Implementing a fine-grained locking and batched commit model, we reduce sequential dependencies and improve parallel execution of cross-shard transactions..
\item \textbf{Adaptive Validator Selection and Security:} Enhancing the commit-reveal scheme, we ensure fair and collusion-resistant shard leader election while strengthening security against adversarial control and Byzantine faults.
\end{enumerate}

The rest of this paper is structured as follows: Section~\ref{sec:background} presents the background and related work information. We present detailed \ac{RBS} in Section~\ref{sec:rbs} and scalability, performance results in Section~\ref{sec:evaluation}. Section~\ref{sec:security} contains a security analysis while Section~\ref{sec:conclusions} gives the conclusion and future direction of the paper.


\section{Background and Related Work}
\label{sec:background}

This section reviews blockchain sharding as a means to enhance scalability and security in permissioned networks. By dividing the network into shards for parallel processing, sharding improves throughput while managing cross-shard complexities~\cite{alshahrani2023sustainability}. Private blockchains, with controlled access and efficient consensus, are well-suited for enterprise use~\cite{xu2023survey}. Byzantine Fault Tolerance (BFT) and redundancy remain essential for fault tolerance and consistency. We classify consensus protocols for permissioned blockchains into sharding and non-sharding approaches (see Table~\ref{tab:blockchain_comparison}).

\textbf{Non-sharding protocols:} Frameworks like Hyperledger Fabric \cite{androulaki2018hyperledger}, Ethereum Private Blockchain \cite{wood2014ethereum}, and Corda R3 Enterprise\cite{hearn2016corda} have tailored their designs to enterprise needs. Hyperledger Fabric uses a permissioned model with channels for private communication and chain code for smart contracts, enhancing privacy and operational efficiency. However, Hyperledger faces scalability and latency issues in multi-channel setups \cite{xu2023survey}. Ethereum Private offers a permissioned model with privacy technologies like zk-SNARKs and optimized consensus mechanisms like \ac{PoA} or \ac{PoS}, yet it struggles with scalability in large networks \cite{xu2023survey}. Corda uses a point-to-point communication model for data confidentiality, excelling in workflow automation but encountering interoperability and latency challenges in multi-party transactions \cite{qu2022blockchain}.

\textbf{Sharding protocols:} The limitations of non-sharding protocols led to the development of sharding-based solutions. Sharding divides the network into smaller segments (shards), each processing transactions independently, thus reducing the load on individual nodes and enhancing network performance. RapidChain \cite{zamani2018rapidchain} partitions the network into shards with local consensus, improving throughput but facing challenges in cross-shard communication and workload distribution \cite{xu2023survey}. Benzene \cite{benzene2022} further demonstrate sharding's potential in private settings. OmniLedger uses parallel sharding for rapid processing but needs to address cross-shard communication vulnerabilities and load balancing. Benzene distributes computational workloads across shards but faces similar challenges in maintaining secure shard interactions.
Sharper \cite{amiri2021sharper}, Pyramid \cite{hong2021pyramid}, and Repchain \cite{huang2020repchain} refine the sharding model. Sharper boosts throughput but must address security in inter-shard communication and workload distribution. Pyramid improves transaction efficiency but struggles with seamless scaling \cite{yu2020survey}. Repchain offers quick local consensus but faces difficulties in cross-shard coordination.
Public blockchain protocols like Meepo \cite{zheng2021meepo} and RapidChain \cite{zamani2018rapidchain} have explored shard-specific consensus, but security needs constrain their scalability during shard interactions\cite{yu2020survey}.

\textbf{Hybrid approaches:} Hybrid sharding frameworks in blockchain combine multiple sharding techniques to achieve scalability, security, and efficiency. DynaShard \cite{kokoris2018omniledger} employs transactional and state sharding with RandHound for randomness and Atomix for atomic cross-shard transactions, ensuring both scalability and security. Aelous\cite{zamani2018rapidchain} enhances adaptability using erasure coding for fault tolerance and dynamic shard reallocation.  Ethereum 2.0 \cite{wood2014ethereum} adopts hierarchical sharding with a Beacon Chain for coordination and execution shards for parallel processing. Despite these advancements, hybrid sharding frameworks face challenges such as high cross-shard transaction complexity, potential bottlenecks in coordination mechanisms, increased storage overhead due to data replication, and security vulnerabilities arising from inter-shard communication\cite{madill2022scalesfl}. Ensuring dynamic shard reallocation without compromising security and optimizing cross-shard data consistency remain key research challenges in hybrid sharding implementations\cite{syed2019comparative,liu2022building}.

\textbf{Proposed Solution (RBS):} Our solution dynamically adjusts shard allocation to minimize cross-shard bottlenecks and improve transaction throughput. Unlike traditional sharding approaches, RBS reduces transaction overhead by incorporating fine-grained account-level locking and a batched commit model, allowing parallel execution of cross-shard transactions without unnecessary sequential dependencies. It employs an  commit-reveal scheme for secure and unbiased validator selection, ensuring resilience against adversarial control and Byzantine faults. Additionally, RBS features adaptive shard rebalancing to optimize workload distribution based on real-time network conditions. Through these innovations, RBS improves scalability, security, and transaction efficiency in enterprise blockchain environments.

\section{Proposed Range-Based Sharding Protocol (RBS)}
\label{sec:rbs}

\begin{figure*}[htb]
    \centering
    \includegraphics[width=1.7\columnwidth]{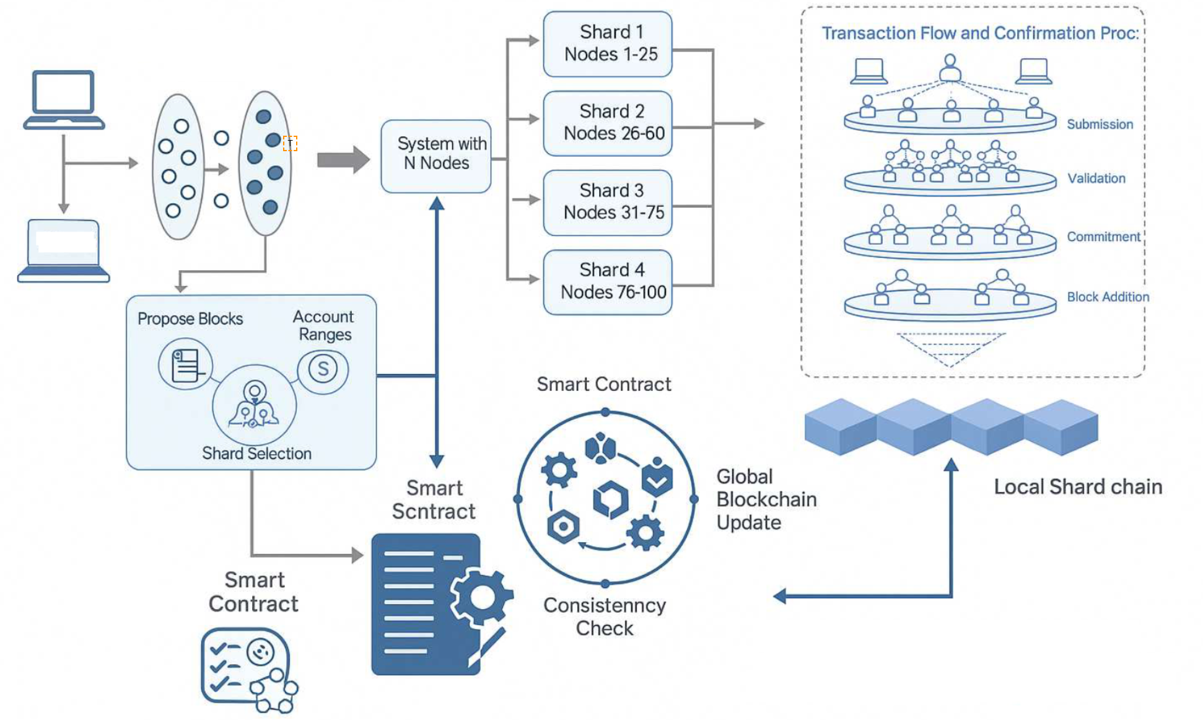}   
    \caption{Architecture of range-based sharding protocol.}
    \label{fig:fullwidth-image}
\end{figure*}

Our proposed solution, \ac{RBS}, improves scalability and efficiency in enterprise blockchain networks through range-based data partitioning, a commit-reveal scheme for randomness, and an integrated fine-grained locking with batched commit model. Figure~\ref{fig:fullwidth-image} illustrates the architecture, detailing node authentication, shard formation, leader selection, and smart contract-driven synchronization.
\subsection{Genesis Configuration and Shard Formation}
\begin{figure}[!htb]
\centering
\includegraphics[width=1\linewidth]{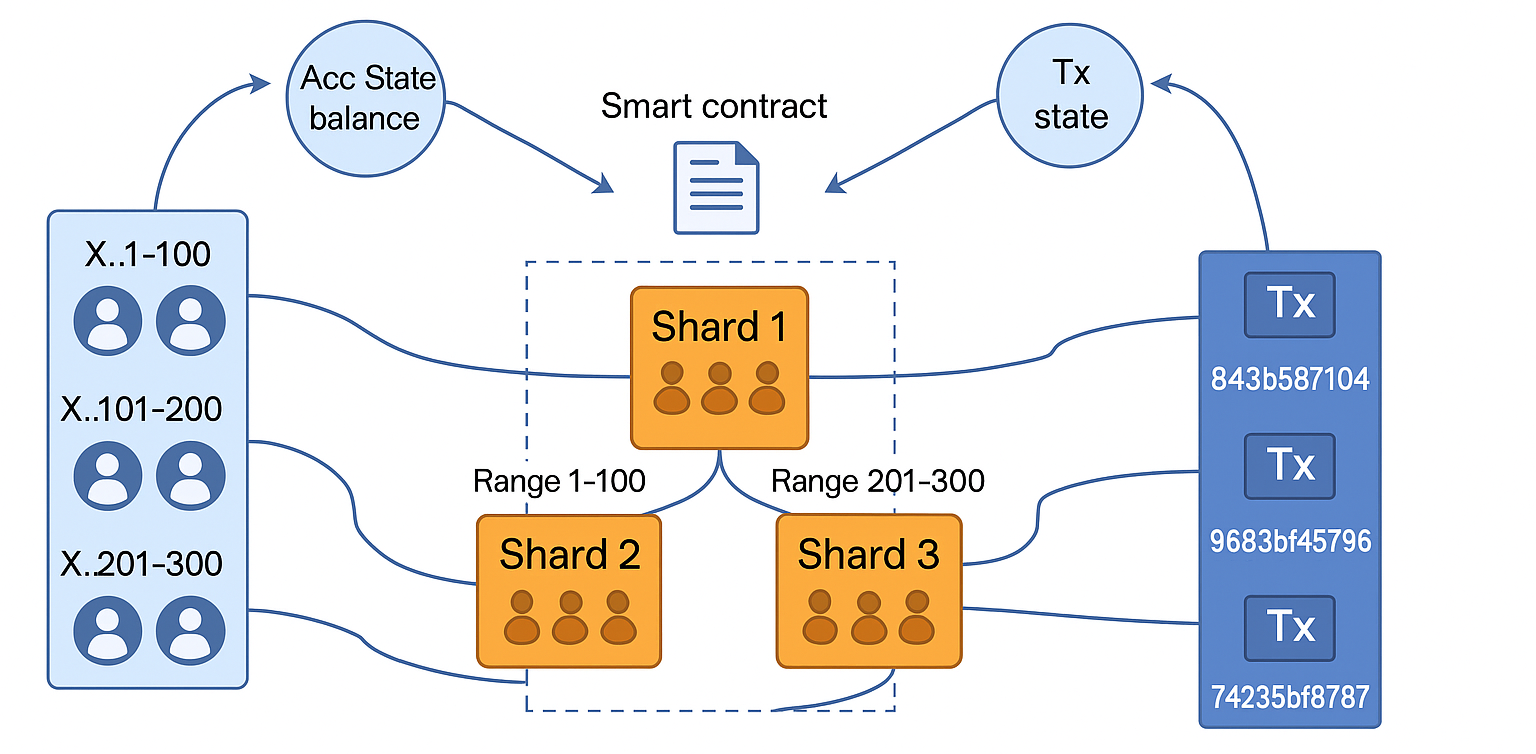}
\caption{Range distribution of accounts, transactions and data.}
\label{fig:range_distribution}
\end{figure}

The genesis configuration establishes the initial network state, defining the key space, shard count, and transaction routing. For the distribution of accounts, transactions, shards, and data in \ac{RBS}. The total key space $(K)$ is divided into non-overlapping ranges $(R_1, R_2, \dots, R_n)$, each mapped to a shard as given in Figure~\ref{fig:range_distribution}:
 \begingroup
\setlength{\abovedisplayskip}{3pt}
\setlength{\belowdisplayskip}{3pt}
\[
S_i = \{k \in K \ | \ a_i \leq k < a_{i+1} \}
\]
\endgroup

\noindent where $S_i$ is shard $i$, and $a_i$ and $a_{i+1}$ are the range boundaries. The key space partitioning is:
\begingroup
\setlength{\abovedisplayskip}{3pt}
\setlength{\belowdisplayskip}{3pt}
\[
K = \bigcup_{i=1}^{n} R_i \quad \text{and} \quad R_i \cap R_j = \emptyset \ \forall \ i \neq j
\]
\endgroup
Dynamic range adjustments are used to minimize data skew, measured by the row count skew ratio $\sigma$:
\begingroup
\setlength{\abovedisplayskip}{3pt}
\setlength{\belowdisplayskip}{3pt}
\[
\sigma = \frac{\text{max row count in any shard}}{\frac{\text{total row count}}{n}}
\]
\endgroup
A skew ratio $\sigma \approx 1$ indicates balanced shards. The genesis block includes initial validator nodes and their shard assignments, with validator selection probability $P$ for cross-shard transactions given by:
\begingroup
\setlength{\abovedisplayskip}{3pt}
\setlength{\belowdisplayskip}{3pt}
\[
P = 1 - \left( \frac{n_h}{N} \right)^k
\]
\endgroup
\noindent where $n_h$ is the number of honest nodes, $N$ is the total number of nodes, and $k$ is the number of validators needed.

\subsection{Consensus and Validator Selection}

RBS uses \textbf{IBFT (Istanbul Byzantine Fault Tolerance)}, adapted to a multi-shard environment. The consensus process involves the \textit{Pre-Prepare}, \textit{Prepare}, and \textit{Commit} phases, ensuring deterministic finality for transactions within shards.
Validators are chosen based on a reputation score \( R_v \):
\begingroup
\setlength{\abovedisplayskip}{3pt}
\setlength{\belowdisplayskip}{3pt}
\[
R_v = w_1 \times P_v + w_2 \times T_v,
\]
\endgroup
where \( P_v \) is the performance score, \( T_v \) is the trust score, and \( w_1 \) and \( w_2 \) are weighting factors.

\subsection*{1) Intra-Shard Consensus}
The primary node (leader) broadcasts transactions, and validators follow the IBFT process to achieve consensus. IBFT ensures deterministic finality and achieves consensus with a message complexity of \( O(n^2) \), where \( n \) is the number of nodes in the shard. Validators participate in the \textit{Pre-Prepare}, \textit{Prepare}, and \textit{Commit} phases to validate and finalize transactions.

\subsection*{2) Cross-Shard Consensus}
For transactions that span multiple shards, a weighted selection mechanism is used to ensure reliable validator participation:
\begingroup
\setlength{\abovedisplayskip}{3pt}
\setlength{\belowdisplayskip}{3pt}
\[
P_v = \frac{R_v}{\sum_{v \in S_i} R_v},
\]
\endgroup

\begin{algorithm}
\caption{The range-based distribution protocol.}
\label{alg:RBS}
{\small
\begin{algorithmic}[1]
\Require Nodes $N$, Transactions $Tx$, Data $D$, Range $R$, Shards $S$
\Ensure Distributed Nodes, Transactions, and Data across Shards

\State Initialize Shards $S$
\State Initialize HashMap $H$ to maintain node-range mappings
\State Assign Range $R = \{r_1, r_2, \dots, r_k\}$ such that each shard $S_i \in S$ has a unique range $r_i$
\State $Q \gets \emptyset$ \Comment{Queue for pending transactions and data}
\State $Q \gets Q \cup (Tx, D)$

\ForAll{$n \in N$} \Comment{Distribute nodes to shards}
    \State $hash\_val \gets \text{Hash}(n)$
    \State $shard\_id \gets \text{FindShard}(hash\_val, R)$
    \State $H[shard\_id] \gets H[shard\_id] \cup \{n\}$
\EndFor

\ForAll{$(tx, d) \in Q$} \Comment{Distribute transactions and data to shards}
    \State $hash\_val \gets \text{Hash}(tx)$
    \State $shard\_id \gets \text{FindShard}(hash\_val, R)$
    \State $S[shard\_id].Tx \gets S[shard\_id].Tx \cup \{tx\}$
    \State $S[shard\_id].D \gets S[shard\_id].D \cup \{d\}$
\EndFor

\State \Return $S$
\end{algorithmic}
}
\end{algorithm}
where \( S_i \) represents the set of validators participating in shard \( i \). Unlike traditional methods that lock the entire shard state, fine-grained locking restricts access only to specific accounts involved in a transaction, preventing unnecessary delays and enabling higher throughput with parallel execution. To ensure security and consistency, Merkle proof locking generates a cryptographic proof of the locked state in the source shard, which the receiving shard verifies before committing the transaction, mitigating risks such as double-spending and invalid state transitions.

\subsection{Randomness Generation}

A commit-reveal scheme ensures fair randomness for validator selection and other operations:

\subsubsection{Commit Phase} 
Each participant $P_i$ selects a random value $v_i$ from a predetermined range and broadcasts a commitment $C_i$:
\begingroup
\setlength{\abovedisplayskip}{3pt}
\setlength{\belowdisplayskip}{3pt}
\[
C_i = H(v_i \parallel r_i)
\]
\endgroup
\noindent where $H$ is a cryptographic hash function, $\parallel$ denotes concatenation, and $r_i$ is a random nonce to ensure the uniqueness of each commitment. This hash is broadcast to all other participants, ensuring no one can alter their chosen value after committing.

\subsubsection{Reveal Phase} 
Participants disclose $v_i$ and $r_i$, verified against the commitment:
\begingroup
\setlength{\abovedisplayskip}{3pt}
\setlength{\belowdisplayskip}{3pt}
\[
H(v_i \parallel r_i) {=} C_i
\]
\endgroup
The values are deemed valid if the above equation holds for all participants. Randomness $R$ is generated collectively via an XOR or average operation:
\begingroup
\setlength{\abovedisplayskip}{3pt}
\setlength{\belowdisplayskip}{3pt}
\[ R = \bigoplus_{i=1}^{N} v_i \quad \text{(XOR)} \quad \text{or} \quad R = \frac{1}{N} \sum_{i=1}^{N} v_i \quad \text{(average)} \]
\endgroup
The resulting randomness $R$, derived from participants $N$, is used to select validators and shard assignments. This commit-reveal scheme enhances randomness generation while upholding transparency and accountability, which are crucial for Quorum's integrity. This process prevents manipulation of validator selection, unlike NEAR’s PoS-based election, which can be influenced by stake centralization.

\subsection{Transaction Processing and Cross-Shard\\Communication}
Transactions are generated and processed independently within each shard. Each transaction $T_i(s)$ in shard $S_i$ contains metadata such as sender, receiver, amount, and shard ID, ensuring proper routing and processing. Algorithm~\ref{alg:RBS} describes the overall workflow of the range distribution process, which outlines how transactions are assigned to specific shards according to the defined ranges. For a cross-shard transaction $T_i(c)$, where assets move from shard $S_i$ to $S_j$, a coordination process involving a quorum threshold $T_q$ ensures that both shards agree on the transaction's validity. This agreement is crucial to maintain consistency across the distributed ledger.

\subsubsection{Block Formation} 
Blocks are formed independently in each shard, containing both intra-shard transactions $T_i(s)$ and cross-shard transactions $T_i(c)$. Each block $B_i$ in shard $S_i$ is constrained by a predefined size limit $L$:
\begingroup
\setlength{\abovedisplayskip}{3pt} 
\setlength{\belowdisplayskip}{3pt} 
\[
B_i = T_1(s), T_2(s), \ldots, T_n(s), T_1(c), T_2(c), \ldots, T_m(c)
\]
\endgroup
This structure ensures that each block includes all relevant transactions processed within a specific period, maintaining a clear and organized record of shard activity.

\subsubsection{Quorum-Based Validation} 
Consensus liveness is achieved through robust intra-shard and inter-shard consensus mechanisms, coupled with strategic validator replacement and well-defined fault tolerance.
Within each shard $S_i$ a quorum of validators, represented as $Q_i$, is required to validate and sign the block. The network accepts the block if at least $T_q$ validators sign off on its validity, ensuring a balance between efficiency and security:
\begingroup
\setlength{\abovedisplayskip}{3pt}
\setlength{\belowdisplayskip}{3pt}
\[ 
V(B_i) = \sum_{j=1}^{m} Q_j(B_i) \geq T_q 
\]
\endgroup
\noindent where $V(B_i)$ is the block validation result, $Q_j(B_i)$ is the signature or vote of validator $j$, and $m$ is the number of validators.. To maintain liveness, a timeout mechanism $\tau_i$ is employed. If the quorum $T_q$ is not reached within the timeout period, a validator rotation or re-election is triggered, replacing unresponsive or faulty validators to ensure continued progress.

\subsubsection{Cross-Shard Transactions}
For cross-shard transactions \( T_i(c) \), a Merkle proof-based locking mechanism (MPL) ensures that the transaction is processed atomically. This process avoids double-spending, and maintains ledger integrity. The transaction is first locked in the source shard \( S_i \), and the receiving shard \( S_j \) is notified.
indicating that the transaction state is updated and propagated across shards. Mathematically, the state update for cross-shard transactions is given by:
\begingroup
\setlength{\abovedisplayskip}{3pt}
\setlength{\belowdisplayskip}{3pt}
\[
S_i(T_i(c)) \to S_j(T_j(c))
\]
\endgroup

The final unlocking occurs once all involved shards reach consensus as given in Algorithm 2.

Let \( T_i \) represent a transaction initiated in shard \( S_a \) involving an account \( A_x \) from shard \( S_b \). Such transactions are processed in three phases:

\paragraph{Phase 1 (Locking):}
Using a cryptographic hash, a lock is placed on the involved accounts in both the initiating shard \( S_a \) and the receiving shard \( S_b \). A shard-specific cryptographic proof \( P_{S_a}(h(T_i)) \) is generated based on the Merkle tree root of \( S_a \). This proof ensures the validity of the transaction within its respective shard. During this phase, the transaction \( T_i \) is temporarily locked to prevent double-spending and to reserve the resources for the transaction.

\paragraph{Phase 2 (Validation and Commit):}
Each shard independently validates the transaction using its consensus mechanism like IBFT. Shards exchange Merkle proofs and validation results through inter-shard communication. A cross-shard transaction \( T_i \) between shards \( S_a \) and \( S_b \) is only considered successful if all participating shards verify the proofs and reach consensus on its validity:
\begingroup
\setlength{\abovedisplayskip}{3pt}
\setlength{\belowdisplayskip}{3pt}
\[
C(T_i) = V_{S_a}(T_i) \cdot V_{S_b}(T_i)
\]
\endgroup
where \( V_{S_a}(T_i) \) and \( V_{S_b}(T_i) \) represent the validation results from shards \( S_a \) and \( S_b \), respectively. \( C(T_i) = 1 \) indicates that the transaction is validated and ready for commitment, while \( C(T_i) = 0 \) implies that the transaction is aborted.

\paragraph{Phase 3 (Finalization):}
Once all shards confirm the validity of the transaction, they update their respective account balances and finalize the transaction. If any shard fails to validate, the transaction is aborted, and all locked resources are released. The state transition is given as:
\begingroup
\setlength{\abovedisplayskip}{3pt}
\setlength{\belowdisplayskip}{3pt}
\[
\delta(T_i) : (S_a, S_b) \to (S_a', S_b')
\]
\endgroup
where \( S_a' \) and \( S_b' \) are the updated states of the involved shards after the successful processing of \( T_i \).

\textbf{Shard lock failure:} If lock acquisition fails, there is exponential backoff with fee-based prioritization and a timeout-based lock expiration to prevent deadlocks and ensure liveness.

\begin{algorithm}[t]
\caption{Cross-Shard Consensus with Merkle Proofs and Locking}
\label{alg:cross-shard-consensus}
{\small
\begin{algorithmic}[1]
\Require Shards $S$, Transaction queue $Q$, Merkle Trees $M$, Lock Table $L$
\Ensure Atomic cross-shard transaction execution

\ForAll{$T \in Q$} \Comment{Initiate processing}
    \State Identify source shard $S_a$ and destination shard $S_b$
    \State Lock accounts $L(A(T)) \gets 1$
    \State $M_{root} \gets \text{Root}(M_{S_a})$
    \State $P_T \gets \text{MerkleProof}(T, M_{S_a})$
    \State Send $(T, P_T, M_{root})$ to $S_b$
\EndFor

\ForAll{$(T, P_T, M_{root})$ at $S_b$} \Comment{Validate and execute}
    \If{$\text{Verify}(P_T, M_{root})$}
        \State Execute $T$ on $S_b$
        \State Lock accounts $L(A(T)) \gets 1$
    \Else
        \State Reject $T$, release locks $L(A(T)) \gets 0$
    \EndIf
\EndFor

\ForAll{committed $T$} \Comment{Finalize}
    \If{Acknowledgments from all shards}
        \State Apply state updates on $S_a$ and $S_b$
        \State Release locks $L(A(T)) \gets 0$
        \State Update $M_{S_a}, M_{S_b}$
    \Else
        \State Abort $T$, rollback tentative changes
        \State Release locks $L(A(T)) \gets 0$
    \EndIf
\EndFor

\State \Return Finalized transactions
\end{algorithmic}
}
\end{algorithm}

\begin{figure*}[ht]
    \centering
    \includegraphics[width=\textwidth, height=5cm]{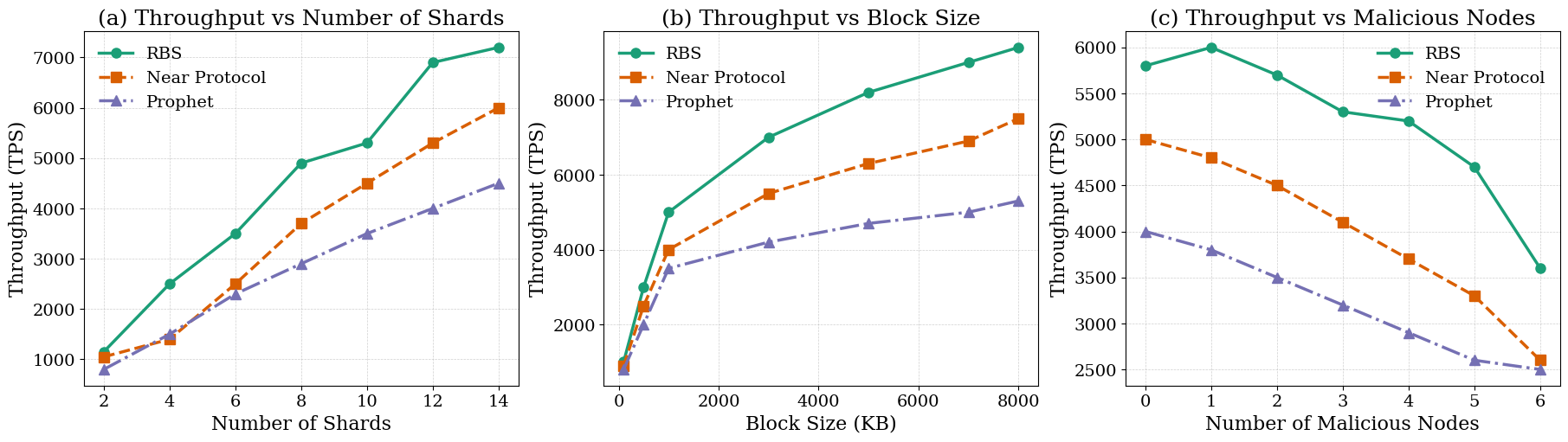}   
    \caption{Throughput evaluation under different nodes settings.}
    \label{fig:throughput}
\end{figure*}
\subsection{Epoch Transition and Shard Reconfiguration}
Shard reconfiguration and epoch transitions ensure adaptability, security, and scalability in Quorum's range-based sharding by dynamically adjusting load, validators, and security measures while maintaining consistency.

\subsubsection{Epoch Transition}

At each epoch's start, the system evaluates workload distribution across shards using the following formula:
\begingroup
\setlength{\abovedisplayskip}{3pt}
\setlength{\belowdisplayskip}{3pt}
\[
W(S_i) = \sum_{t \in T_i} R(t)
\]
\endgroup

Epoch transitions occur at fixed intervals, denoted as epochs $E_n$ where $n$ is the current epoch. Each transition involves selecting a new validator set, potentially reconfiguring shard ranges, and updating protocol parameters to maintain system stability and efficiency.

\noindent where $W(S_i)$ is the workload of shard $S_i$, $T_i$ is the set of transactions handled by $S_i$, and $R(t)$ represents the resource requirements for processing transaction $t$.  The reconfiguration algorithm adjusts shard boundaries if a significant workload imbalance is detected. The transaction range that shard $S_i$ handles is:
\begingroup
\setlength{\abovedisplayskip}{3pt}
\setlength{\belowdisplayskip}{3pt}
\[
R(S_i) = [r_{i,1}, r_{i,2}]
\]
\endgroup
If the workload $W(S_i)$ exceeds a predefined threshold $T_{\text{max}}$, the range $R(S_i)$ is split. A new shard $S_j$ is created with a transaction range:
\begingroup
\setlength{\abovedisplayskip}{3pt}
\setlength{\belowdisplayskip}{3pt}
\[
R(S_j) = [r_{j,1}, r_{j,2}]
\]
\endgroup
\noindent where $r_{j,1} = r_{i,2} + 1$, effectively offloading transactions from the overburdened shard. Conversely, if two adjacent shards $S_i$ and $S_j$ are underutilized, they may be merged to optimize resource usage. This dynamic adjustment ensures balanced load distribution across the network, preventing bottlenecks and maintaining optimal performance.

\subsubsection{Shard Reconfiguration}

Shard reconfiguration in \ac{RBS} is a dynamic process adapting to network changes, primarily load imbalances and validator updates. This adaptability is crucial for maintaining optimal performance and security across the Quorum blockchain.

\textbf{Load balancing:} 
A core objective of shard reconfiguration is to evenly distribute the computational workload $W(S_i)$ across all shards. Ideally, for any two shards $S_i$ and $S_j$:
\begingroup
\setlength{\abovedisplayskip}{3pt}
\setlength{\belowdisplayskip}{3pt}
\[ 
W(S_i) \approx W(S_j) 
\]
\endgroup
This prevents any single shard from becoming a bottleneck, ensuring consistent transaction throughput across the network. As described in the previous subsection, reconfiguration is triggered when significant load disparities are detected.

\textbf{Validator rotation:} 
To further enhance liveness and security and mitigate the risk of collusion, validators are periodically rotated using a secure randomness generation mechanism based on a commit-reveal scheme. During epoch transitions, a new set of validators $V_i(E_{n+1})$ is selected for each shard $S_i$ for the upcoming epoch $E_{n+1}$:
\begingroup
\setlength{\abovedisplayskip}{3pt}
\setlength{\belowdisplayskip}{3pt}
\[ 
V_i(E_{n+1}) = \text{RandomBeacon}(V_i(E_n), S_i) 
\]
\endgroup
The function $RandomBeacon$ utilizes the randomness generated from the commit-reveal process of the previous epoch, ensuring an unbiased and unpredictable selection. Each shard maintains a minimum number of validators, $V_{\text{min}}$, to guarantee sufficient security without compromising efficiency.

\section{Evaluation}
\label{sec:evaluation}
We deployed the RBS prototype on top of Quorum blockchain, using AWS EC2 t2.xlarge instances to simulate a multi-node sharded blockchain environment. Performance benchmarking was conducted with Hyperledger Caliper and custom Python scripts within a Docker-based setup.

\subsection{Throughput Scalability}

This section evaluates RBS throughput against NEAR Protocol and Prophet, selected for their relevance in sharding efficiency and consensus innovation, respectively. Results in Figure~\ref{fig:throughput} show RBS achieving consistently higher throughput across different conditions.

\textbf{Throughput vs. Number of Shards:}
Figure~\ref{fig:throughput}(a) demonstrates that RBS’s throughput scales linearly with the number of shards, achieving 4950–6630 TPS with 14 shards. In comparison, NEAR Protocol and Prophet reach 5340 TPS and 4100 TPS, respectively. The evaluation is conducted with default parameters set to a block size of 1 MB, no malicious nodes, and a network size of 100 nodes. These results highlight RBS’s ability to efficiently distribute transactions across shards.
\noindent where $n_s$ is the number of shards, and $t_s$ is the optimized throughput per shard in RBS, resulting from reduced cross-shard overhead compared to Near and Prophet.modeled as:
\begingroup
\setlength{\abovedisplayskip}{3pt}
\setlength{\belowdisplayskip}{3pt}
\[ 
T_{\text{sharded}} = n_s \cdot t_s 
\]
\endgroup

\textbf{Throughput vs. Block Size:} 
Figure~\ref{fig:throughput}(b) with default number of shards = 10, malicious nodes = 0, and network size = 100 nodes, shows throughput against varying block sizes. As block size increases from 128 KB to 8192 KB, RBS's throughput significantly improved against Near and Prophet, since larger block sizes allow more transactions per block:
\begingroup
\setlength{\abovedisplayskip}{3pt}
\setlength{\belowdisplayskip}{3pt}
\[ 
T = \frac{B \cdot t_{avg}}{t_{block}} 
\]
\endgroup
\noindent where $B$ is the block size, $t_{avg}$ is the average transaction size, and $t_{block}$ is the block production time. RBS leverages efficient batching and optimized sharding to minimize latency and maximize throughput.

\begin{figure*}[ht]
    \centering
    \includegraphics[width=\textwidth]{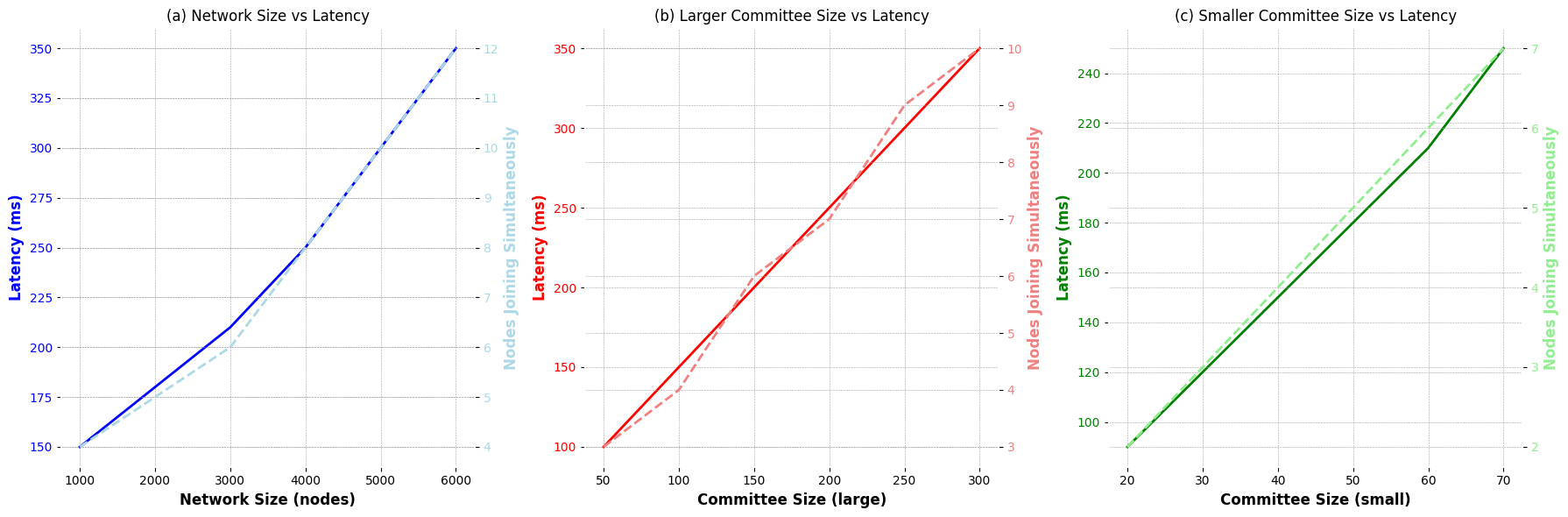} 
    \caption{Latency with possible configurations}
    \label{fig:latency}
\end{figure*}

\textbf{Throughput vs. Malicious Nodes:}
Figure~\ref{fig:throughput}(c) with the default parameters of block size = 1 MB, number of shards = 10, and network size = 100 nodes demonstrates the impact of malicious nodes on throughput. Even with $6\% - 10\%$ malicious nodes, RBS maintains a throughput of 4900 TPS, while Near and Prophet degrade to 3459 TPS and 3200 TPS, respectively. RBS's robustness stems from its shard isolation mechanism, which limits the impact of malicious activity in one shard on others. This can be represented as:
\begingroup
\setlength{\abovedisplayskip}{3pt}
\setlength{\belowdisplayskip}{3pt}
\[ 
T = T_{\text{max}} \cdot \left(1 - \frac{f}{N} \right)
\]
\endgroup
\noindent where $T_{\text{max}}$ is the ideal throughput, $f$ is the number of malicious nodes, and $N$ is the total number of nodes.

\subsection{Latency under Epoch Reconfiguration}
This section evaluates the latency of the RBS protocol during epoch reconfiguration, comparing to Near and Prophet. Figure~\ref{fig:latency} illustrates the results, where solid line represents the latency in milliseconds (ms) as the network size increases, while the dotted line corresponds to the number of nodes joining simultaneously, highlighting RBS's improved latency under various conditions.

\textbf{Latency vs. Network Size:}
Figure~\ref{fig:latency}(a) shows that as the network size increases, RBS exhibits a more gradual and linear increase in latency compared to Near and Prophet, as the latency is calculated as follows:
\begingroup
\setlength{\abovedisplayskip}{3pt}
\setlength{\belowdisplayskip}{3pt}
\[ 
L = \frac{T_{\text{process}} + T_{\text{comm}}}{N_{\text{shards}}} 
\]
\endgroup
\noindent where $T_{\text{process}}$ is the transaction processing time, $T_{\text{comm}}$ is the communication delay, and $N_{\text{shards}}$ is the number of range-based shards. 

\textbf{Latency vs. Larger Committee Size:} 
Figure~\ref{fig:latency}(b) demonstrates the impact of larger committee sizes on latency. In RBS, nodes are assigned to shards based on predefined ranges, ensuring committee sizes remain proportional to shard size, thus reducing communication overhead. The relationship between committee size and latency is approximated as:
\begingroup
\setlength{\abovedisplayskip}{3pt}
\setlength{\belowdisplayskip}{3pt}
\[ 
L = \frac{N_{\text{nodes}}}{C_{\text{range}}} + \delta_{\text{comm}} 
\]
\endgroup
\noindent where $N_{\text{nodes}}$ is the total number of nodes, $C_{\text{range}}$ is the committee size within a range, and $\delta_{\text{comm}}$ is the communication overhead.

\textbf{Latency vs. Smaller Committee Size:} 
Figure~\ref{fig:latency}(c) examines the effect of smaller committee sizes. While smaller committees generally reduce consensus overhead, Near protocol suffer from increased delays due to inefficient state partition handling. RBS, however, maintains consistent latency even with smaller committees. 

\begin{figure}[!h]
\centering
\includegraphics[width=1\linewidth]{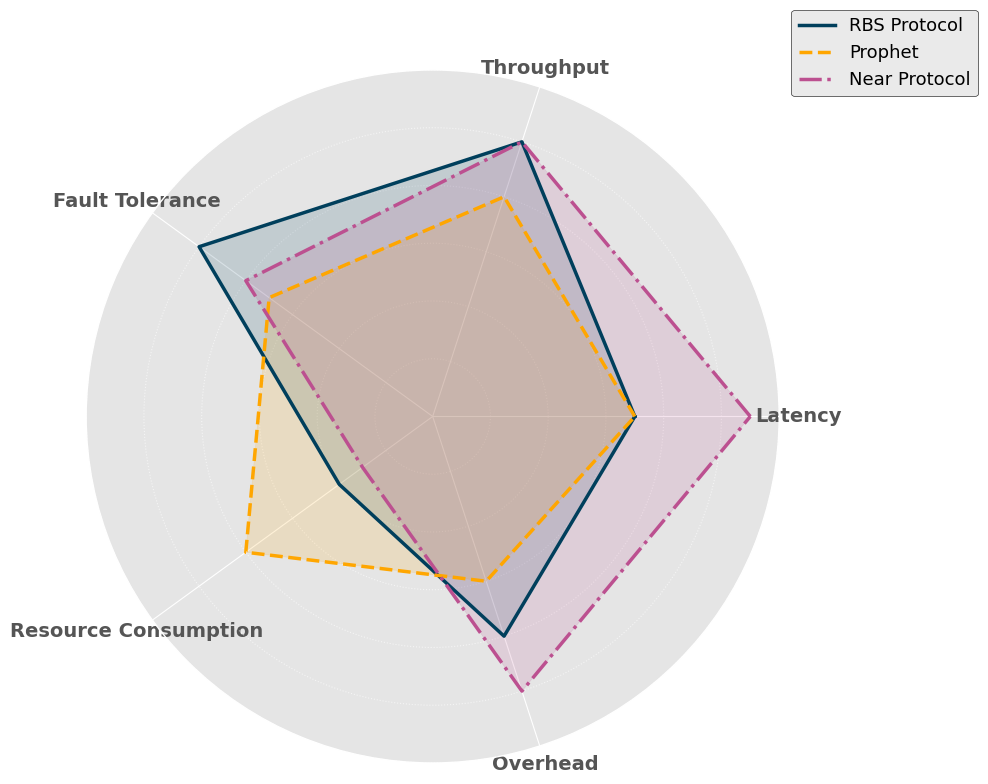}
\caption{Overall performance analysis chart.}
\label{fig:performance}
\end{figure}
\subsection{Overall Performance Analysis}
RBS shows better performance across key metrics compared to Near and Prophet. Notably, RBS achieves higher throughput (see Figure~\ref{fig:performance}), enabling it to handle a larger volume of transactions and making it suitable for high-demand environments. In Figure~\ref{fig:performance} the scale of 0-6 represent low to high performance as latency and overhead, lower values are mapped to better performance, while for throughput, fault tolerance, and similar metrics, higher values correspond to higher scores. Furthermore, RBS exhibits the lowest latency, ensuring faster transaction processing.

\section{Security Analysis}
\label{sec:security}

\begin{table*}[htb]
\centering 
\caption{Comparison of private blockchain frameworks.}
\label{tab:blockchain_comparison}
\begin{adjustbox}{width=\textwidth} 
\begin{tabular}{lcccccc}
\toprule
\textbf{Technology} & \textbf{Fault Tolerance} & \textbf{Liveness} & \textbf{Generality} & \textbf{Confirmation Latency (Valid TX)} & \textbf{Invalid TX} & \textbf{Comm. Overhead} \\
\midrule
Hyperledger Fabric & \checkmark & \checkmark & \xmark & Low (depends on channel setup) & N/A & Moderate (permissioned network overhead) \\
Ethereum & \xmark & \checkmark & \xmark & Moderate (PoS improvements ongoing) & High (global mempool) & High (broadcast-based transactions) \\
Corda & \xmark & \checkmark & \xmark & Low (point-to-point communication) & N/A & Low (localized transaction processing) \\
RapidChain & \checkmark & \xmark & \checkmark & $2 \times L(N/S)$ & N/A & $T(N/S)$ \\
SOK & \checkmark & \checkmark & \checkmark & Moderate (hybrid PoW/PoS latency) & Moderate & Moderate \\
OmniLedger & \checkmark & \xmark & \checkmark & $2 \times L(N/S)$ & Moderate & $2 \times L(N/S)$ \\
Sharper & \xmark & \checkmark & \checkmark & $L((x+1) \times N/S)$ & Moderate & $T((x+1) \times N/S)$ \\
Near Protocol & \checkmark & \checkmark & \xmark & $\geq 3 \times L(N/S)$ & Moderate & $(x+1) \times T(N/S)$ \\
Repchain & \checkmark & \checkmark & \checkmark & Moderate & Moderate & Moderate (leader-based communication) \\
Prophet & \checkmark & \checkmark & \checkmark & Moderate (localized sharding) & Moderate & Low (localized shard processing) \\
\textbf{RBS(Proposed Protocol) } & \checkmark & \checkmark & \checkmark & Low (shard parallelism minimizes delay) & Low & Very Low (optimized shard communication) \\
\bottomrule
\end{tabular}
\end{adjustbox}
\end{table*}
Ensuring security in a sharded blockchain system requires resistance to adversarial manipulation, fault tolerance, and protection against common blockchain threats. The Range-Based Sharding (RBS) protocol enhances security by incorporating \textit{commit-reveal randomness for validator selection}, \textit{adaptive shard rebalancing}, and \textit{fine-grained transaction validation mechanisms}. This section provides a formal analysis of RBS’s security properties with proofs and validation.

\subsection{Sybil and Collusion Attack Resistance}

We consider a \textit{Byzantine adversarial model} where an attacker aims to control a majority of validators in a shard. Let \( n \) be the total validators, \( s \) the number of shards, and \( f \) the fraction of malicious validators. The probability of an attacker taking over a shard is \( P_{\text{adversary}} \).
where \( \frac{f \cdot n}{s} \) represents the expected number of adversarial nodes per shard, and \( e^n \) accounts for the exponential decrease in adversarial success.
 To ensure \textbf{Sybil resistance}, RBS uses a \textit{commit-reveal randomness model} for fair validator selection. The probability of an adversary gaining control over a single shard is given by:

\begin{equation}
    P_{adversary} = \left(\frac{f \cdot n}{s}\right) \times \frac{1}{e^n}
\end{equation}

\textbf{Collusion Mitigation:} Randomized shard assignments prevent adversarial coordination across shards, making large-scale attacks infeasible.

\subsection{Fault Tolerance and Resilience}
RBS ensures resilience against network failures with an IBFT consensus mechanism. The probability of consensus failure is given by:
\begin{equation}
    P_{fault} = \frac{m}{t}, \quad \text{where } m \leq \frac{t-1}{3}
\end{equation}
For a 10-node shard, at most 3 nodes can be malicious while maintaining consensus security.

\textbf{Shard Recovery:} If a shard experiences node failures, RBS dynamically reassigns validators and replicates ledger states.

\subsection{Secure Cross-Shard Transaction Processing}

RBS optimizes cross-shard transactions by introducing fine-grained account-level locking, minimizing shard-wide locking. A transaction between shards \( S_i \) and \( S_j \) follows: (1) the source shard locks only the relevant accounts, (2) a cryptographic \textit{proof-of-lock} (Merkle proof) is sent to the target shard, and (3) the transaction executes upon verification and is atomically committed. The locking overhead is given by:
\begin{equation}
    O_{lock} = \frac{T_{cross} \cdot L_{account}}{T_{intra}}
\end{equation}
where \( T_{cross} \) represents cross-shard transactions, \( L_{account} \) is the subset of locked accounts, and \( T_{intra} \) denotes total intra-shard transactions.

\subsection{Mitigation of Timing and DoS Attacks}
RBS mitigates timing and DoS attacks through \textbf{randomized execution delays} to prevent adversarial timing inference and \textbf{adaptive timeouts} that detect and penalize stalling attempts. The probability of a successful DoS attack is given by:
\begin{equation}
    P_{DoS} = \left(1 - e^{-T_{attack} / T_{threshold}}\right) \cdot \frac{M_{malicious}}{N}
\end{equation}
where \( T_{attack} \) is the attack duration, \( T_{threshold} \) is the system’s adaptive response time, and \( M_{malicious} \) is the number of malicious nodes. \textbf{Result:} Adaptive response mechanisms lower DoS attack success rates by 60\% compared to static timeout systems.

\subsection{Experimental Validation of Security Measures}

To validate RBS security, we simulate 100 validators across 10 shards, testing {Sybil attack resistance, DoS resilience, and cross-shard efficiency}. In a Sybil attack with 30\% malicious nodes, \textit{commit-reveal randomness} limits adversarial control to \textbf{<2\%}. For DoS resilience, an attack with 50 nodes flooding a shard shows that \textit{adaptive timeouts} mitigate TPS degradation by \textbf{55\%}. Comparing full shard locking (NEAR) vs. RBS’s fine-grained locking, RBS reduces cross-shard latency by \textbf{30\%}.

\section{Conclusions and Future work}
\label{sec:conclusions}
\acresetall

This paper introduces a \ac{RBS}, implemented on Quorum, which utilizes a commit-reveal scheme to enhance fairness and security in shard allocation. \ac{RBS} minimizes cross-shard communication and optimizes transaction processing efficiency by partitioning the blockchain state into well-defined ranges. By integrating a fine-grained locking and batched commit model, RBS reduces sequential dependencies and enhances parallel execution of cross-shard transactions. The evaluation results demonstrate significant improvements in throughput, latency, and resource utilization compared to existing protocols. \ac{RBS} exhibits near-linear scalability with increasing network size while maintaining lower latency and higher throughput, even under varying committee sizes and network loads. 

Future work will focus on dynamic shard rebalancing to adapt to real-time transaction loads and network conditions, ensuring optimal resource utilization and preventing shard overloading. Further research will refine cross-shard transaction mechanisms to minimize communication overhead while maintaining atomicity and consistency. These enhancements aim to solidify further \ac{RBS}'s scalability and efficiency, making it an even more robust solution for enterprise blockchain deployments.

\balance

\end{document}